\documentclass[aps,amsmath,onecolumn,floatfix,superscriptaddress,prl,footinbib,dvipdfmx,10pt]{revtex4-2}
\usepackage{physics}
\usepackage{graphicx} 
\usepackage{dcolumn} 
\usepackage{bm} 
\usepackage{braket} 
\usepackage{ulem} 
\usepackage[usenames,dvipsnames]{color}
\usepackage[
colorlinks=true, citecolor=blue, urlcolor=blue, linkcolor=blue,
setpagesize=false, bookmarks=false
]{hyperref}
\usepackage{siunitx}
\usepackage{mhchem}
\usepackage{comment}
\usepackage{mathrsfs}

\begin{document}

\title{Floquet Theory and Ultrafast Control of Magnetism} 

\author{Masahiro Sato}
\affiliation{Department of Physics, Ibaraki University, Mito, Ibaraki 310-8512, Japan}

\begin{abstract}
The development of laser science and technology have stimulated the study of condensed matter physics, especially, dynamical or non-equilibrium nature in solids. 
The laser technique in terahertz (THz) regime, whose photon energy is comparable to those of typical collective modes in solids such as magnetic excitations, phonons, etc., has remarkably proceeded in the last decade. 
Theoretical tools for non-equilibrium states have also progressed. Thanks to these backgrounds, magneto-optics, 
especially, the study of controlling magnetism with laser, now enters a new stage. For such controls, 
Floquet engineering is a key concept, which means the method of controlling static properties of targets 
with high-frequency external fields like laser. 
I review the theoretical foundation of Floquet engineering and its application to magnetic insulators.
Basic magnetic quantities such as magnetization, spin chirality, and spin current are shown to be controlled 
with intense THz laser or wave.
\end{abstract}

\maketitle
\tableofcontents

\section{Introduction}\label{sec:1}
Laser science and technology have continuously developed in the last decades. 
The application of laser to solids is being one of the hottest topics in condensed-matter physics. 
If we apply intense laser to materials, their quantum states quickly change into a non-equilibrium one and 
we can observe non-equilibrium or relaxation dynamics, nonlinear responses to the intense AC field, 
ultrafast change of physical quantities, etc. 
In recent years, the significant development of terahertz (THz) laser 
science in the range of $\sim$0.1-10 THz~\cite{Hirori11,Mukai16,Cavalleri17,Misawa13} (THz=$10^{12}$Hz) 
has accelerated the study of ultrafast control of magnetism with THz laser 
because its photon energy is comparable to the energy of magnetic excitations, especially, 
those of antiferromagnets~\cite{Kimel18}. 
The maximum intensity of currently available THz laser has attained the electric-field amplitude 1-10 [MV/cm] 
which corresponds to a few Tesla of the AC magnetic field amplitude. 
In addition to THz laser science, the magnetic resonance study with THz or gigahertz (GHz) waves (10GHz $-$ 1.0THz) 
have been long investigated~\cite{Slichter,Nojiri12} (GHz=$10^{9}$Hz). 
The control of magnetism with laser or electromagnetic wave~\cite{Kimel18,Kirilyuk10} 
has also gathered much attention as a large branch of spintronics~\cite{SpinCurrent12}. 

Recently, several theoretical methods for non-equilibrium systems have also developed, and 
they have gradually become widespread in broad fields of condensed-matter physics.  
Non-equilibrium Green's function method~\cite{Zagoskin,Stefanucci,Haug,Kamenev,Aoki}, 
approaches based on quantum master equation~\cite{Breuer,Alicki,Weiss,Breuer00,Ikeda19-2}, 
and Floquet theory~\cite{Bukov,Eckardt15,Eckardt17,Oka19} are representative of them. 
These techniques have a high potential to capture different aspects of laser-driven non-equilibrium dynamics. 
In fact, with such methods, I and collaborators have theoretically explored/proposed several ways 
of controlling physical properties of materials, especially, focusing on 
magnetic systems~\cite{Takayoshi14,Sato16,Sato14,Fujita17,Fujita18,Fujita19,Higashikawa,Ishizuka19,Ikeda19}. 

Among studies for laser-driven phenomena, the concept ``Floquet engineering'' has 
been an important keyword and provided us various research directions. 
This terminology stands for controlling physical (especially static) properties of target systems 
by applying an AC field whose frequency (photon energy) is much higher than the energy scale of the systems. 
Excitations or quasi particles of focused systems cannot be directly coupled to such a high-frequency AC field, 
but it is known that static or low-frequency properties of the systems can be changed 
through the nonlinear effects of the AC field. 
This effect is theoretically formulated by Floquet theorem and 
related techniques developed in recent years~\cite{Bukov,Eckardt15,Eckardt17,Oka19}.

In this Chapter, I would like to review the theoretical basis of Floquet engineering 
and its application to simple, realistic magnetic systems. I will explain that basic magnetic quantities 
such as magnetization, spin chirality, and spin current can be controlled 
by application of intense THz laser or wave to magnets.  

\section{Floquet Engineering}\label{sec:2}
This section is devoted to the explanation about the theoretical basis of 
Floquet engineering~\cite{Bukov,Eckardt15,Eckardt17,Oka19}. 
As I mentioned, Floquet engineering means creating non-equilibrium states with desirable (static) 
physical properties by periodically driving (i.e., by applying an external AC field to) a material. 
This concept stems from Floquet theorem, and so I start from the explanation about the theorem.  
Then I will derive the Floquet effective Hamiltonian through so-called high-frequency (Floquet-Magnus) expansion. 
The effective Hamiltonian is the most important instrument in Floquet engineering 
from its conceptual viewpoint, and it describes slow dynamics of the driven system.   
Finally, I will state some remarks on the physical meaning of Floquet Hamiltonian.

As one will see soon later, the Floquet theory based on Floquet theorem assumes that 
(i) the external AC field is treated as a classical number (not operator) in the Hamiltonian considered, 
and (ii) the driven system is decoupled to any environment, i.e., we consider ``isolated'' quantum systems.
The assumption (i) might be sometimes justified. 
For instance, if the intensity of applied laser is large enough, 
its AC electric and magnetic fields can be approximated 
by classical external fields. 
The condition (ii) is too ideal, especially, if we consider usual materials such as solids, liquids, and gases. 
However, as I will explain below, the ideal conditions (i) and (ii) enable us to reveal
a clear, simple picture of Floquet engineering. 
The engineering in dissipative driven systems~\cite{Breuer,Aoki,Ikeda19-2} 
is a front line of the non-equilibrium physics.

\subsection{Floquet Theorem}\label{sec:2-1}
Floquet theorem is an old mathematical result for a class of linear differential equations with a time periodic term, 
but in the field of physics, it has been recognized as a theorem for Schr\o dinger equation (i.e., equation of motion)
for periodically-driven quantum systems. Following this convention, 
I will prove Floquet theorem for quantum systems in this subsection. 
Hereafter, I will often use the unit of $\hbar=1$.

The statement of Floquet theorem is as follows. 
We start from the time-dependent Schr\o dinger equation for a periodically driven quantum system
\begin{align}
\label{eq:Schrodinger}
i\frac{\partial}{\partial t}\Psi(t)=\hat H(t)\Psi(t). 
\end{align}
Here, we assume that the Hamiltonian $\hat H(t)$ is time periodic, $\hat H(t+T)=\hat H(t)$, 
where $T=2\pi/\omega$ is the period and $\omega$ is the (angular) frequency. 
If we focus on a system driven by laser or electromagnetic wave, $\omega$ is the laser frequency. 
For this driven quantum system, the theorem shows that the solution of Eq.~(\ref{eq:Schrodinger}) is given by
\begin{align}
\label{eq:Floquet}
\Psi(t)=\exp(-i\epsilon t) \Phi(t),
\end{align}
where the ``wave function'' $\Phi(t)$ is a periodic one satisfying $\Phi(t+T)=\Phi(t)$ and 
the real number $\epsilon$ is called Floquet quasi energy. Namely, Floquet theorem states that 
the solution of Schr\o dinger equation for a periodically driven system is given by the product of 
a plane wave $e^{-i\epsilon t}$ (i.e., solution of vacuum) and a periodic function $\Phi(t)$. 
In this sense, Floquet theorem can be viewed 
as the time version of Bloch theorem (See, e.g., Refs.~\cite{Ashcroft,Grosso}) for spatially-periodic quantum systems.

Let us prove the above statement. We define the one-cycle time-evolution operator as 
\begin{align}
\label{eq:Time}
\hat U(t+T,t) \equiv {\cal T} \Big[\exp\Big(-\frac{i}{\hbar} \int_t^{t+T}d\tau \hat H(\tau)\Big)\Big],
\end{align}
where the symbol ${\cal T}$ denotes time-ordered product. 
From the periodicity of the Hamiltonian, $\hat U(t+T,t)=\hat U(t+(n+1)T,t+nT)$ for $n\in Z$. 
If $\hat U(t+T,t)$ acts on $i\partial_t\Psi(t)=\hat H(t)\Psi(t)$ from the left in both sides, 
we obtain $i\partial_t\Psi(t+T)=\hat H(t+T)\Psi(t+T)=\hat H(t)\Psi(t+T)$.
That is, $\Psi(t)$ and $\Psi(t+T)$ both satisfy the same Schr\o dinger equation. 
Therefore, for a solution $\Psi(t)$, there always exists another solution $\Psi(t+T)$ which is proportional to $\Psi(t)$: 
$\Psi(t+T)=c_T(t)\Psi(t)$. The coefficient $c_T(t)$ is regarded as the eigenvalue of $\hat U(t+T,t)$. 
Namely, we have
\begin{align}
\label{eq:Phasefactor}
\hat U(t+T,t) \Psi(t) &= c_T(t) \Psi(t). 
\end{align}
Since $\hat U(t+T,t)$ is unitary, its eigenvalue $c_T(t)$ is just a phase factor, $|c_T(t)|=1$.  

First, I show that $c_T(t)$ is $t$-independent ($c_T(t)=c_T$), i.e., a time-evolved state 
$\Psi(t')=\hat U(t',t)\Psi(t)$ with any $t'$ is also the eigenstate of $\hat U(t'+T,t')$ 
with the same eigenvalue $c_T$. To this end, let us make $\hat U(t,t')\hat U(t',t)=\hat 1$ 
act on Eq.~(\ref{eq:Phasefactor}) from the left side. The left hand side is calculated as 
$\hat U(t,t')\hat U(t',t)\hat U(t+T,t) \Psi(t)=\hat U(t,t')\hat U(t'+T,t+T)\hat U(t+T,t) \Psi(t)
=\hat U(t,t')\hat U(t'+T,t) \Psi(t)=\hat U(t,t')\hat U(t'+T,t')\Psi(t')$, 
where we have used the relation $\hat U(t',t)=\hat U(t'+T,t+T)$. The right hand side 
is done as $\hat U(t,t')\hat U(t',t) c_T(t)\Psi(t)=c_T(t)\hat U(t,t')\Psi(t')$. 
If we further multiply both the sides by $\hat U(t',t)$, then we obtain 
\begin{align}
\label{eq:Phasefactor2}
\hat U(t'+T,t') \Psi(t') &= c_T(t) \Psi(t'). 
\end{align}
Equations (\ref{eq:Phasefactor}) and (\ref{eq:Phasefactor2}) reveal that $c_T(t)$ is independent of $t$.

Next, let us turn to the remaining part, the proof of $c_T=e^{-i\epsilon T}$.  
Thanks to the periodicity of $\hat H(t+T)=\hat H(t)$, the Hamiltonian and 
the one-cycle time-evolution operator $\hat U(t+T,t)$ commute with each other: 
$[\hat U(t+T,t),\hat H(t)]=0$, which means that 
$\hat H(t)$ and $\hat U(t+T,t)$ can be simultaneously diagonalized. 
Moreover, the operator $\hat U(t+T,t)$ follows the relation,
\begin{align}
\label{eq:Time2}
\hat U(t+(m+n)T,t+mT)\hat U(t+mT,t)&=\hat U(t+mT,t)\hat U(t+(m+n)T,t+mT)\nonumber\\
&=\hat U(t+T,t)^{m+n}\nonumber\\
&=\hat U(t+(m+n)T,t).
\end{align}
This means that the eigenvalue $c_T$ of $\hat U(t+T,t)$ satisfies $c_{nT}c_{mT}=c_{mT}c_{nT}=c_{(m+n)T}=c_T^{m+n}$. 
Therefore, to realize this equality, $c_T$ has to be exponential, that is, $c_T=e^{-i\epsilon T}$. 
Combining this eigenvalue and the nature of simultaneous diagonalization, 
we can say that the solution $\Psi(t)$ follows
\begin{align}
\label{eq:Floquet2}
\Psi(t+T)=e^{-i\epsilon T}\Psi(t).
\end{align}
If we introduce $\Phi(t)=e^{i\epsilon t}\Psi(t)$, $\Phi(t)$ is shown to be a periodic function as follows:
\begin{align}
\label{eq:Floquet3}
\Phi(t+T)=e^{i\epsilon (t+T)}\Psi(t+T)=e^{i\epsilon (t+T)}e^{-i\epsilon T}\Psi(t)=e^{i\epsilon t}\Psi(t)=\Phi(t). 
\end{align}
We thereby arrive at Eq.~(\ref{eq:Floquet}).

\subsection{Discretized Fourier Transformation and Matrix Form of Schr\o dinger Equation}\label{sec:2-2}
Bloch theorem for solid crystals leads to their electron band structure~\cite{Ashcroft,Grosso}. 
Following a similar manner, I here show that a generalized eigenvalue problem appears 
by applying Floquet theorem to periodically-driven systems. 
It tells us a basic physical picture of the driven systems.

The periodicity of $\hat H(t+T)=\hat H(t)$ and $\Phi(t+T)=\Phi(t)$ indicate that 
$\hat H(t)$ and $\Phi(t)$ can both be Fourier-transformed along time direction in a discretized way. 
We define their Fourier transforms as 
\begin{align}
\label{eq:Fourier}
\hat H(t)=\sum_{m\in Z} e^{-im\omega t} \hat H_m, & \,\,\,\,\,\,\,\,\,\,\,\,\,\,\,
 \Phi(t)=\sum_{m\in Z} e^{-im\omega t} \Phi_m. 
\end{align}
The inverse transformation is given by
\begin{align}
\label{eq:Fourier2}
\hat H_m = T^{-1} \int_0^T dt \,\, e^{im\omega t} \hat H(t), & \,\,\,\,\,\,\,\,\,\,\,\,\,\,\, 
\Phi_m = T^{-1} \int_0^T dt \,\,e^{im\omega t} \Phi(t). 
\end{align}
Substituting these and Eq.~(\ref{eq:Floquet}) into the Schr\o dinger equation, 
we obtain the following generalized eigenvalue problem for $\{\hat H_m\}$ and $\{\Phi_m\}$: 
\begin{align}
\label{eq:GeneralEigen}
\sum_{n\in Z}(\hat H_{m+n}-m\omega \delta_{m,n})\Phi_n &= \epsilon \Phi_m. 
\end{align}
Therefore, with the set of Hamiltonians $\{\hat H_m\}$, we can compute the wave functions $\{\Phi_m\}$ 
and the corresponding quasi energy $\epsilon$ (i.e., eigenvalue) in principle. 
Since the real number $\epsilon$ is introduced in the form of $e^{-i\epsilon T}$,  
its physical relevant range is ristricted in the ``Brillouin'' zone 
$-\frac{\pi\hbar}{T}=-\frac{\hbar\omega}{2}\leq \epsilon< \frac{\hbar\omega}{2}=\frac{\pi\hbar}{T}$ 
like crystal momentum $k$ in solids. 
Due to this periodicity, $\epsilon$ is called {\it quasi} energy. 
We emphasize that Eq.~(\ref{eq:GeneralEigen}) does not explicitly depend on time $t$, 
and in that sense, it may be called a {\it static} eigenvalue equation.

In order to more deeply understand this static equation, let us re-write it in a matrix form. 
Equation~(\ref{eq:GeneralEigen}) can be expressed as 
\begin{equation}
\label{eq:Matrix}
\left(\begin{matrix}
\ddots &  \ddots                 & \ddots               & \ddots  &                 &                               &             \\
\cdots & \hat H_0-2\hbar \omega & \hat H_1                    & \hat H_2    & \hat H_3                    & \cdots                &              \\
\cdots & \hat H_{-1}                    & \hat H_0-\hbar\omega & \hat H_1    & \hat H_2                    & \hat H_3                    & \cdots    \\
\cdots & \hat H_{-2}                    & \hat H_{-1}                 & \hat H_0    & \hat H_1                    & \hat H_2                    & \cdots    \\
\cdots & \hat H_{-3}                    & \hat H_{-2}                 & \hat H_{-1} & \hat H_0+\hbar\omega & \hat H_1                    & \cdots    \\
         & \cdots                   & \hat H_{-3}                 & \hat H_{-2} & \hat H_{-1}                 & \hat H_0+2\hbar\omega &  \cdots   \\
        &                             &                          & \ddots & \ddots                 & \ddots                 & \ddots \\
\end{matrix}\right)
\left(\begin{matrix}
\vdots \\
\Phi_2 \\
\Phi_1 \\
\Phi_0 \\
\Phi_{-1} \\
\Phi_{-2} \\
\vdots \\
\end{matrix}\right)
=\epsilon
\left(\begin{matrix}
\vdots \\
\Phi_2 \\
\Phi_1 \\
\Phi_0 \\
\Phi_{-1} \\
\Phi_{-2} \\
\vdots \\
\end{matrix}\right).
\end{equation}
\begin{figure}[t]
\begin{center}
\includegraphics[scale=.55]{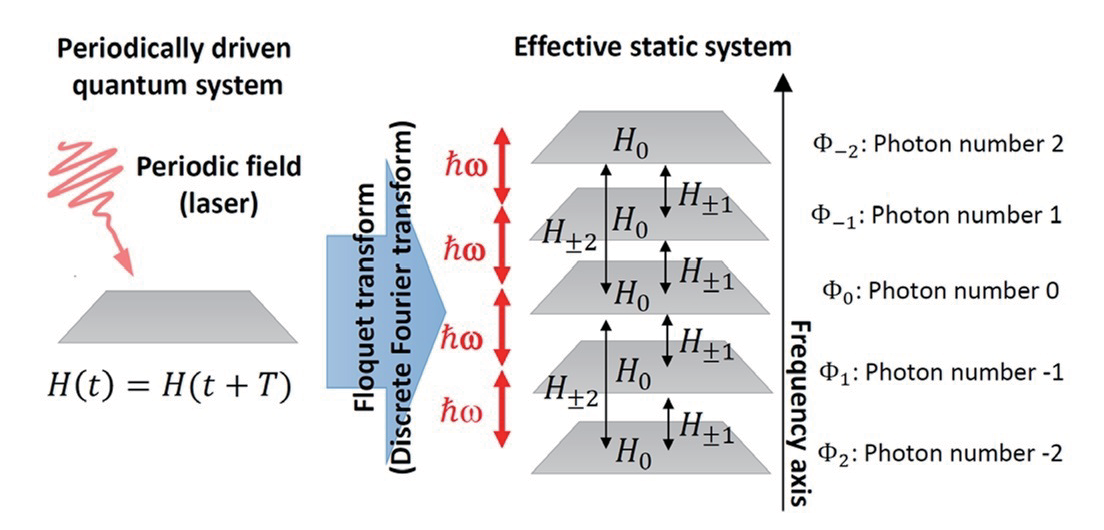}
\caption{Images of a periodically driven quantum system and 
the mapped static system through Floquet (Fourier) transform.}
\label{fig:FloquetMap}
\end{center}
\end{figure}
Here we have explicitly restored the symbol $\hbar$. 
The wave function $\Phi_n$ on each line is a vector living in a Hilbert space 
whose width is the same as that of the original driven system. 
An image of this matrix representation is given in Fig.~\ref{fig:FloquetMap}. 
We can obtain an intuitive picture of the eigenvalue equation~(\ref{eq:Matrix}), i.e., 
Eq.~(\ref{eq:GeneralEigen}) if the external AC field is viewed as laser with photon energy $\hbar\omega$. 
From the diagonal components of Eq.~(\ref{eq:Matrix}), we find that 
the energy decreases (increases) by $\hbar\omega$ whenever one goes up (down) the lines one by one.  
Therefore, $\{\cdots,\Phi_{-2},\Phi_{-1},\Phi_0,\Phi_1,\Phi_2,\cdots\}$ may be viewed as 
wave functions in subspaces with different photon numbers $\{\cdots,2,1,0,-1,-2,\cdots\}$, respectively. 
The diagonal part of the subspace with photon number $n$ is given by $\hat H_0+n\hbar\omega$, 
where $\hat H_0=T^{-1}\int_0^{T}dt \hat H(t)$ is the time averaged Hamiltonian. 
On the other hand, from the off-diagonal part of Eq.~(\ref{eq:Matrix}), we see that 
$\hat H_{\pm n}$ ($n\neq 0$) connects two subspaces whose difference of photon numbers is $n$.  

In summary, Floquet theorem enables us to exactly map a non-equilibrium quantum system 
with periodic driving to a ``static'' eigenvalue problem, Eq.~(\ref{eq:GeneralEigen}) or (\ref{eq:Matrix}).   
As well known, various theoretical tools have been developed to analyze static many-body (eigenvalue) problems 
thanks to the long history of equilibrium statistical and condensed-matter physics (See e.g., Refs.~\cite{AGD,Mahan}). 
In this sense, Floquet theory makes a periodically-driven system transformed to an easier problem. 
However, we note that instead of the emergent static nature, a new index $n$ (photon number) appears, 
and it means that the ``spatial'' dimension increases by unity if the direction of $n$ is viewed as a new spatial axis. 

\subsection{Floquet-Magnus Expansion and Floquet Hamiltonian}\label{sec:2-3}
It is known that one can compute an effective static Hamiltonian 
(called Floquet Hamiltonian)~\cite{Shirley,Samba} acting on the original space of 
the Schr\o dinger equation~(\ref{eq:Schrodinger}) 
from the eigenvalue equation~(\ref{eq:GeneralEigen}) or (\ref{eq:Matrix}).
The approximation is based on the power series expansion of $1/\omega$. 
There are several sorts of the expansions and they are related with each other~\cite{Mikami16}. 
In general, they are called Floquet-Magnus (high-frequency) expansion.

Here, I explain one of the expansion methods which is based on degenerate perturbation theory, 
by making use of the matrix representation~(\ref{eq:Matrix}) and Fig.~\ref{fig:FloquetMap}. 
Since the bra-ket notation is generally useful for the perturbation calculation, 
we re-write wave functions $\Phi_n$ and $\Psi(t)$ as $|\Phi_n\rangle$ and $|\Psi(t)\rangle$, respectively, 
in this subsection. 
We consider the case where the photon energy $\hbar\omega$ is sufficiently larger than other energy scales 
(i.e., all the eigenvalues of $\{\hat H_{n}\}$). 
In this condition, the ``distance'' $\hbar\omega$ between neighboring subspaces is large enough and 
therefore the averaged Hamiltonian $\hat H_0$ and Fourier components $\hat H_{\pm n}$ $(n\neq 0)$
connecting different subspaces may be viewed as the perturbation with respect to the the diagonal photon energies 
${\rm diag}(\cdots,2\hbar\omega,\hbar\omega,,0,-\hbar\omega,-2\hbar\omega,\cdots)$.

Below, I will construct the effective Hamiltonian for a subspace with a fixed photon number. 
Hereafter, we call the subspace with photon number $n$ as $n$th subspace, and 
ignore the unperturbed photon energies $n\hbar\omega$ since they are just constants. 
In fact, owing to periodicity of the quasi energy $\epsilon$, we arrive at the same effective Hamiltonian 
in each subspace with photon number $n$, expect for the constant $n\hbar\omega$. 
We define the orthonormal basis $\{|\Phi_{-n}\rangle_j\}$ ($j=1,2,\cdots,d$) for $n$th subspace, 
where $d$ is the dimension of the subspace. Let us introduce 
the photon-number state $|n\rangle$ that is convenient for the perturbation calculation.  
With it, the projection operator $\hat P_{-n}$ to the $n$th subspace may be expressed as 
$\hat P_{-n}=\sum_{j=1}^d|\Phi_{-n}\rangle_j {}_j\langle\Phi_{-n}|=|n\rangle\langle n|\times
\sum_{j=1}^d|\Phi\rangle_j {}_j\langle\Phi|$. The basis $|\Phi\rangle_j$ is common to all the subspaces 
and in that sense, two projection operators $\hat P_{-n}$ and $\hat P_{-m}$ ($n\neq m$) 
are equal to each other, $\hat P_{-n}\sim \hat P_{-m}$, except for their photon states. 
Similarly, the averaged Hamiltonian $\hat H_0$ in the diagonal part of Eq.~(\ref{eq:Matrix}) 
should be represented as $\hat H_0 |n\rangle\langle n|$, and $\hat H_m$($m\neq 0$) connecting 
$n$th and $(n+m)$th subspaces as $\hat H_m |n+m\rangle\langle n|$. One can perform 
the perturbation calculation utilizing these instruments.

The first-order Hamiltonian is equivalent to putting the perturbation term between the projection operator. 
Therefore, we obtain 
\begin{align}
\label{eq:1stPerturbation}
\hat H^{(1)} \equiv \hat P_{-n}\hat H_{0}\hat P_{-n}=\hat H_0.
\end{align}
Fourier components $\hat H_{n\neq0}$ all disappear due to the sandwich of $\hat P_{-n}$. 
We also note that $\sum_{j=1}^d|\Phi\rangle_j {}_j\langle\Phi|$ is unity when we focus on a subspace 
with a fixed photon number. 
Following the standard formula of the second-order perturbation, 
we write down the second-order term as 
\begin{align}
\label{eq:2ndPerturbation_1}
\hat H^{(2)} &= \hat P_{-n}\sum_{m=1}^\infty\Biggl( 
\sum_j \frac{\hat H_{+m}|\Phi_{-(n+m)}\rangle_j {}_j\langle\Phi_{-(n+m)}|\hat H_{-m}}{n\hbar\omega-(n+m)\hbar\omega}
\nonumber\\
&+
\sum_j \frac{\hat H_{-m}|\Phi_{-(n-m)}\rangle_j {}_j\langle\Phi_{-(n-m)}|\hat H_{+m}}{n\hbar\omega-(n-m)\hbar\omega}
\Biggr)\hat P_{-n}
\end{align}
The projection operator $\hat P_{-(n\pm m)}$ in the intermediate process is also viewed as unity 
except for the photon bra-ket $|n\pm m\rangle\langle n\pm m|$. 
Therefore, we arrive at 
\begin{align}
\label{eq:2ndPerturbation_2}
\hat H^{(2)} &= 
\hat P_{-n}\sum_{m=1}^\infty\left( -\frac{\hat H_{+m}\hat H_{-m}}{m\hbar\omega}
+\frac{\hat H_{-m}\hat H_{+m}}{m\hbar\omega}\right)\hat P_{-n} 
= -\sum_{m=1}^\infty \frac{[\hat H_{+m},\,\, \hat H_{-m}]}{m\hbar\omega}.
\end{align}
Namely, the second-order Hamiltonian is given by the sum of commutators between 
two off-diagonal terms $\hat H_{\pm m}$.

Up to the second-order terms, the effective Hamiltonian (Floquet Hamiltonian) is 
\begin{align}
\label{eq:effHam}
\hat H_{\rm eff} &= \hat H_0 -\sum_{m=1}^\infty \frac{[\hat H_{+m},\,\, \hat H_{-m}]}{m\hbar\omega}
+{\cal O}\left((\hbar\omega)^{-2}\right). 
\end{align}
The first term is the time averaged Hamiltonian and often identified with that before applying the external AC field.   
The second and higher-order terms with power of $1/\omega$ emerge only when the AC field is applied.  
The formula~(\ref{eq:effHam}) indicates that these AC-field-driven terms can be controlled by tuning 
kinds, wave forms, and frequency $\omega$ of the AC field. 
That is, the Hamiltonian can be desirably changed by applying a high-frequency AC field in a clever way. 
This statement gives contrast to a stereotype idea that the Hamiltonian is fixed for each material.  
Equation~(\ref{eq:effHam}) clearly shows the basic idea of Floquet engineering. 
In principle, we can compute higher-order terms by continuing the degenerate perturbation theory.
For instance, the third-order term~\cite{Mikami16} is  
\begin{align}
\label{eq:3rdPerturbation}
\hat H^{(3)} &=\sum_{\substack{m=-\infty \\ (m\neq 0)}}^\infty
\frac{[[\hat H_{-m},\hat H_0],\hat H_m]}{2m^2(\hbar\omega)^2}
+\sum_{\substack{m=-\infty \\(m\neq 0)}}^\infty \sum_{\substack{n=-\infty \\(n\neq 0,m)}}^\infty 
\frac{[[\hat H_{-m},\hat H_{m-n}],\hat H_n]}{3mn(\hbar\omega)^2}.
\end{align}

Before ending this subsection, we shortly comment on another expansion 
method~\cite{Mikami16,Bukov,Eckardt15,Eckardt17,Oka19}. 
The eigenvalue of Eq.~(\ref{eq:Matrix}) is the quasi energy $\epsilon$, and (as we already mentioned)
it is defined in the eigenvalue of the one-cycle time evolution operator $\hat U(t+T,t)$ as $e^{-i\epsilon T}$. 
From these facts, the static Floquet Hamiltonian $\hat {H}_{\rm eff}$ may seem to be defined as 
\begin{align}
\label{eq:effHam_2}
\exp(-i\hat H_{\rm eff}T)&=  {\cal T} \Big[\exp\Big(-\frac{i}{\hbar} \int_{0}^{T}d\tau \hat H(\tau)\Big)\Big]
=\hat U(T,0). 
\end{align}
If the phase factor of the right hand side can be expanded with respect to power of $1/\omega$, 
we obtain the series expansion formula of $\hat {H}_{\rm eff}$. 
It is known that if the time evolution operator is assumed to be decomposed in the following form, 
one can compute the expanded form of the effective Hamiltonian: 
\begin{align}
\label{eq:Decompose}
\hat U(t_2,t_1)&= \exp(-i \hat G(t_2))\exp\big[-i\hat {\cal H}\,\,(t_2-t_1)\big]\exp(i \hat G(t_1)).
\end{align}
Here, $\hat G(t)$ is a periodic operator $\hat G(t)=\hat G(t+T)$, and the exponential 
$\exp(\pm i \hat G(t))$ describes the high-frequency fluctuation in one cycle $T$. 
$\hat G(t)$ is called kick operator or micro-motion operator. 
On the other hand, $\hat {\cal H}$ is the time-independent operator 
describing the slow dynamics longer than the period $T$. 
Generally, $\hat {\cal H}$ depends on the initial time $t_1$, but the $t_1$ dependence 
is eliminated under the condition that $\hat G(t)$ satisfies $\int_0^T dt\hat G(t)=0$. 
In this case, the high-frequency expansion of $\hat {\cal H}$ is shown to be equal to that of $\hat {H}_{\rm eff}$ 
in Eq.~(\ref{eq:effHam}). 
Similarly, the kick operator can be expanded as $\hat G(t)=\sum_{n=1}G^{(n)}$.  
The lower-order terms are given by
\begin{align}
\label{eq:1nd2rdkicked}
i\hat G^{(1)}(t) &=-\sum_{\substack{m=-\infty \\ (m\neq 0)}}^\infty \frac{\hat H_m}{m\hbar\omega}e^{-im\omega t}
\\
i\hat G^{(2)}(t)&=\sum_{\substack{m=-\infty \\ (m\neq 0)}}^\infty
\frac{[\hat H_m,\hat H_0]}{m^2(\hbar\omega)^2}e^{-im\omega t}+
\sum_{\substack{m=-\infty \\(m\neq 0)}}^\infty \sum_{\substack{n=-\infty \\(n\neq 0,m)}}^\infty
\frac{[\hat H_n,\hat H_{m-n}]}{2mn(\hbar\omega)^2}e^{-im\omega t}
\end{align}

\subsection{Physical Meaning of Floquet Hamiltonian}\label{sec:2-4}
In the previous subsection, I explained that the static Floquet Hamiltonian 
$\hat H_{\rm eff}$ of Eq.~(\ref{eq:effHam}) is computed through the Floquet-Magnum high-frequency expansion. 
What can we understand from $\hat H_{\rm eff}$? One might expect that thermal equilibrium or 
ground states of the ``Hamiltonian'' $\hat H_{\rm eff}$ is realized by continuously applying an AC field. 
However, such a naive expectation is not correct. 
Below, I will comment on a few important results related with the Floquet Hamiltonian $\hat H_{\rm eff}$, 
focusing mainly on many-body systems.

\begin{description}
\item{[{\bf Short time behavior}]}
When one applies the expansion formulas of $\hat H_{\rm eff}$ and $\hat G(t)$ to a periodically-driven system, 
the Floquet-Magnus expansion should be terminated at a certain order in a practical sense. 
If we focus on a small finite-size system, the expansion is often well-defined. 
On the other hand, it is known that if we consider a wide class of locally-interacting many-body systems, 
their Floquet-Magnus expansion is usually of an asymptotic-expansion type like the perturbation expansion 
of quantum field theories. Let us define a truncated Floquet Hamiltonian as 
$\hat {H}_q \equiv\sum_{k=0}^q\hat {H}^{(k)}$. The following statements about $\hat {H}_q$ 
have been theoretically shown for locally interacting many-body systems~\cite{Kuwahara16,Mori16}. 
(i) There is an optimal number $q=q_0\propto \hbar\omega/G$ 
which is the best truncation order of the Floquet-Magnus expansion. 
Here, $G$ is the typical local energy scale of the system 
including the coupling between the system and AC field. 
(ii) $\hat {H}_{q<q_0}$ behaves as an almost conserved quantity during a short time 
$\tau_1\sim \exp[{\cal O}(\hbar\omega/G)]T$ from the 
beginning of application of AC field. 
(iii) During a short time $\tau_2\ll \tau_1$, the time evolution operator can be approximated as
\begin{align}
\label{eq:FloquetTimeDep}
\hat U(t_2,t_1)&\approx \exp\big[-i\hat {H}_{q<q_0} \,\,(t_2-t_1)\big]. 
\end{align}
as long as we focus on the slow dynamics longer than the period $T$. 
These statements indicate that $\hat {H}_{q<q_0}$ is considered as the Hamiltonian 
for describing the time evolution at least within a short time. 
If we attach a truncated kick operator $G_p(t)=\sum_{n=1}^pG^{(n)}$ to Eq.~(\ref{eq:FloquetTimeDep}), 
a more accurate short-time evolution operator is given by 
\begin{align}
\label{eq:FloquetTimeDep2}
\hat U(t_2,t_1)&\approx \exp(i\hat G_p(t_2))\exp\big[-i\hat {H}_{q<q_0} \,\,(t_2-t_1)\big]\exp(i\hat G_p(t_1)). 
\end{align}
From these arguments, the truncated Floquet Hamiltonian is physically relevant in a short time. 
\\

\item{[{\bf Heating effect}]}
In the present Floquet-theory formalism, we consider isolated (closed) quantum systems decoupled 
to any environment. 
It is believed that if we apply an intense AC field to such a closed system for a long time, 
the system is eventually heated up, except for a class of toy models including integrable systems. 
In fact, this feature is confirmed by theoretical studies for 
some concrete models driven by AC fields~\cite{Lazarides14,D'Alessio14}. 
This result seems to be very natural because if a crystal is irradiated by laser, 
it is generally heated and sometimes evaporates. Namely, everyone knows that 
the heating effect of applied laser is usually unavoidable.  
We also note that the heating effect of AC fields is consistent with the above theoretical results (i)-(iii), 
because the results indicate that a long-time Floquet engineering is generally impossible 
if we consider an isolated system. \\

\item{[{\bf Efficient engineering}]}
The form of the Floquet Hamiltonian (\ref{eq:effHam}) tells us 
that the dimensionless expansion parameter is roughly given by $A/(\hbar\omega)$, 
where $A$ is the typical energy scale of the local interaction between the system and the AC field. 
In order to enhance the accuracy of predictions from the Floquet Hamiltonian truncated at a low order, 
one should increase AC-field frequency $\omega$ or decrease the coupling strength $A$ 
(i.e., tuning the value of $A/(\hbar\omega)$). 
On the other hand, for a small value of $A/(\hbar\omega)$, AC-field driven low-order terms 
such as $\hat H^{(0)}$, $\hat H^{(1)}$, and $\hat H^{(2)}$ are also weak. 
Namely, the change of physical quantities via Floquet engineering is quite small 
for a small $A/(\hbar\omega)$. 
Therefore, one should take a moderate value of $A/(\hbar\omega)$ 
to perform Floquet engineering in an efficient manner (although the Floquet Hamiltonian 
gradually becomes invalid with increase of $A/(\hbar\omega)$).  
The requirement of a large $A$ is quite natural because the Floquet engineering is a typical 
nonlinear phenomenon driven by a strong AC field. 
As I will mention in the next section, it is not easy to increase $A$ if one use laser or electromagnetic wave 
as the driving field of Floquet engineering. 
\end{description}

From these statements, we see that the present Floquet theory can correctly predict the short-time behavior 
when we consider isolated periodically driven many-body systems. 
For a long-time driving, the systems are generally heated and the Floquet theory does not work well. 
However, I emphasize that the forms of low-order terms 
such as $\hat H^{(0)}$, $\hat H^{(1)}$, and $\hat H^{(2)}$ are very important to qualitatively understand 
which kinds of quantities can be controlled by an AC field. 
A more quantitative computation beyond Floquet Hamiltonian (e.g., numerical analysis) 
is often necessary to more accurately predict the values and time evolution of AC-field driven quantities, 
but finding set of possible engineering observables is the most fundamental process 
to propose a novel Floquet engineering.

\section{Laser and Typical Excitations in Solids}\label{sec:3}
Some parts of theoretical proposals for Floquet engineering have been experimentally realized 
by using cold-atom systems~\cite{Eckardt15,Eckardt17} in which one can prepare 
nearly isolated (closed) quantum systems in a few (sub-)seconds. 
It means that Floquet engineering in cold-atom systems has already grown to a certain degree, 
while that in materials still offers various open issues. 
Therefore, now is the time to develop theories and experiments for Floquet engineering in usual materials 
such as solids, liquids, and gases. 
Use of usual materials has the advantage compared with cold atoms. 
For instance, the lifetime of materials may be considered as infinity, and 
thereby one can repeatedly perform Floquet-engineering process with single material. 
Floquet engineering in materials can be done in a table-top manner by applying a suitable AC field 
while that in cold atoms requires application of many sorts of tuned lasers 
together with atom-trapping techniques.

As I mentioned in Introduction, 
I will review a few examples of the Floquet engineering in solid crystals in the next section. 
A typical high-frequency field for solids is electromagnetic wave or laser beams. 
Therefore, the information about currently-available laser and excitations in solids is important. 
Below I discuss it from a quantitative viewpoint.

\begin{figure}[t]
\begin{center}
\includegraphics[scale=.5]{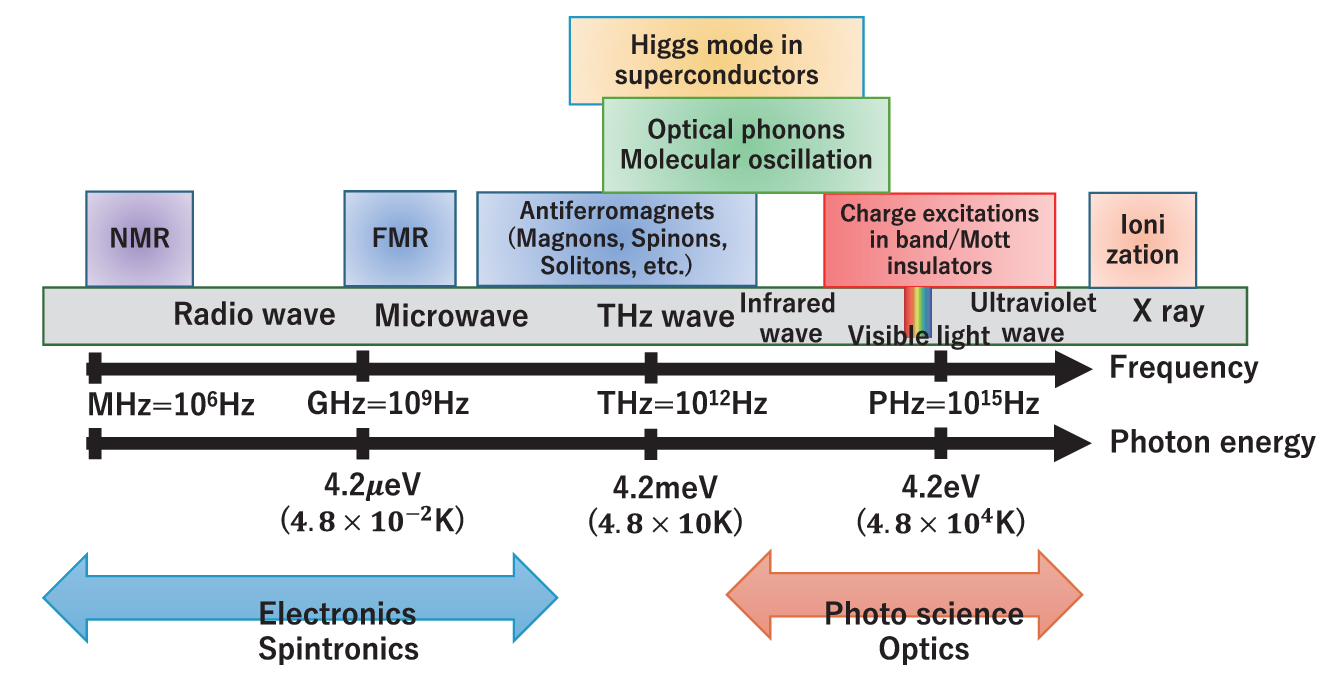}
\caption{Typical excitations in solids and their typical frequency (energy scales) 
in a wide frequency range of electromagnetic wave.}
\label{fig:Excitations}
\end{center}
\end{figure}
Figure~\ref{fig:Excitations} depicts typical excitations (quasi particles) of solids 
in a wide range of laser frequency (photon energy). 
Magnetic (electron spin) excitations in solids are usually distributed from 1 GHz to 10 THz.  
Nuclear spin dynamics is much slower than electron spin motion and 
nuclear magnetic resonance (NMR)~\cite{Slichter} is usually done in a megahertz (MHz=$10^{6}$Hz) range. 
Phonons (lattice vibration) and molecular oscillations are located from 1 THz to infrared wave ($\sim 100$ THz).   
Electron charge excitations are around the visible-light regime ($0.1-10$ petahertz (PHz=$10^{15}$Hz)) 
in both band and Mott insulators. 

Electromagnetic waves and quasi-particle dynamics in MHz and GHz (low-frequency) range 
have been widely used in electronics and spintronics~\cite{SpinCurrent12}. 
In this low-frequency regime, it is difficult to create intense coherent waves like laser beams 
(although there exists maser technique), and 
instead various methods of generating electromagnetic waves have been established. 
For instance, Gunn diode has been often used to create micro waves 
for electron spin resonance (ESR) experiments. 
Techniques based on electromagnetic induction have been often used in NMR. 
On the other hand, the high-frequency waves including infrared, visible and ultraviolet light and 
the corresponding high-energy excitations in solids (various charge excitations) have been 
central instruments in the research of photo science, optics, and magneto-optics~\cite{Kirilyuk10}.  
In the high-frequency regime, several methods of creating laser beams have been established. 

The laser science and technology in THz range (0.1-10THz), which is located between the frequencies 
used in electronics (spintronics) and optics (photo science), have massively proceeded 
in the last decades~\cite{Hirori11,Mukai16,Cavalleri17,Misawa13,Kimel18}. As shown in Fig.~\ref{fig:Excitations}, 
their photon energy is comparable to the collective modes in solids like magnetic excitations, phonons, 
molecular oscillations, Higgs modes in superconductors, etc. Therefore we can now directly control 
these modes in ultrafast ways with intense laser. 
The intensity of THz laser is still weaker than those of higher-frequency laser, 
but methods of controlling both intensity and shape of THz waves have been continuously developed.  
In fact, non-equilibrium magnetic phenomena induced by THz laser or wave 
have been actively investigated in various experimental 
groups~\cite{Mukai16,Baierl16,Lu17,Pimenov06,Takahashi11,Staub14,Sirenko19}. 
As every one well knows, the strongest light-matter coupling is the interaction between 
electric charge and electric field. 
Thus so far charge dynamics in metals and semiconductors has been the central target in photo physics and optics. 
However, thanks to the development of THz laser science, non-equilibrium magnetic phenomena 
induced by direct spin-light couplings have joined in photo science~\cite{Kimel18,Kirilyuk10}.

Usually, laser beams stand for the strong, coherent electromagnetic wave propagating for a long time. 
Such a beam is called continuous wave (CW). In addition to CW, laser pulses 
with a short-time Gaussian envelop curve have been intensively studied, especially, in condensed-matter physics.  
A few cycle, one-cycle and half cycle pulses have been often utilized in the study of photo science.  
Particularly, as I said, since it is difficult to make the intensity of THz laser strong compared to the other lasers,  
the pulse techniques have been often used to create intense THz laser pulse. 
This is very important to generate nonlinear laser-driven phenomena including Floquet engineering 
in the THz range (0.1-10THz). On the other hand, the Floquet theory discussed in Sec.~\ref{sec:2} is 
reliable for systems driven by CW. One thereby carefully apply the Floquet approach 
to theoretically study Floquet engineering in pulse-driven systems.

\begin{table}[!t]
\begin{center}
\caption{Properties of electromagnetic wave (or laser) with 1[MV/cm]. $c$ is speed of light.}
\label{tab:MV}       
\begin{tabular}{p{6cm}p{1cm}p{4cm}p{4cm}}
\hline\noalign{\smallskip}
Electric field of electromagnetic wave & & $E_0=$1\,MV/cm  \\
\noalign{\smallskip}\hline
\noalign{\smallskip}
Magnetic flux density, $B_0=E_0/c$ && 0.33\,T \\
Energy flux, $I$ && 1.3$\times10^{9}$\,W/cm${}^2$=1.3 GW/cm${}^2$\\
\noalign{\smallskip}\hline\noalign{\smallskip}
\end{tabular}
\end{center}
\end{table}
\begin{table}[!t]
\begin{center}
\caption{Electromagnetic waves with frequencies 1GHz and 1THz and related physical quantities. 
Symbols of $k_{\rm B}$, $g=2$, $\mu_{\rm B}$, $c$, $e$, and $a_{\rm B}$ are respectively 
Boltzmann constant, electron g factor in vacuum, Bohr magneton, speed of light, elementary charge, and Bohr radius.}
\label{tab:Units}       
\begin{tabular}{p{6cm}p{1cm}p{2cm}p{2cm}}
\hline\noalign{\smallskip}
Electromagnetic wave &  & 1THz & 1GHz  \\
\noalign{\smallskip}\hline
\noalign{\smallskip}
Frequency, $\omega/(2\pi)$   &  & $10^{12}$\,Hz  & $10^{9}$\,Hz \\
Photon energy, $\hbar\omega$ & & 4.1\,meV& 4.1\,$\mu$eV \\
Temperature, $T=\hbar\omega/k_\text{B}$ &  & 48\,K & 48\,mK \\
Magnetic flux density, $B_0=\hbar\omega/g\mu_\text{B}$ && 36\,T & 36\,mT \\
Electric field (vol. 1), $E_1=cB_0$ &&  107\,MV/cm &  107\,kV/cm\\
Electric field (vol. 2), $E_2=\hbar\omega/(e a_{\rm B})$ && 0.8\,MV/cm  &  0.8\,kV/cm \\
\noalign{\smallskip}\hline\noalign{\smallskip}
\end{tabular}
\end{center}
\end{table}

The laser intensity of 1 [MV/cm] may be viewed as a reference value for observing 
nonlinear photo-induced phenomena, 
and the corresponding amplitude of the AC magnetic field is $\sim$ 0.3 Tesla. See Table~\ref{tab:MV}. 
Let us here reminder that 1 [MV/cm]=0.1 [V/nm], Bohr radius $a_{\rm B}=0.0529$nm, and 
the energy levels of a hydrogen atom are given by $-13.6/n^2$[eV] ($n=1,2,3,\cdots$ is the quantum number). 
Therefore, 1 [MV/cm] of electric field in an atomic size or lattice spacing of crystals ($\sim$1nm) 
is the same as about 10 $\%$ of a typical energy gap between neighboring atomic levels.  
In addition, since typical strength of exchange interactions in magnetic insulators is 10-100 Tesla, 
the magnetic field of 0.3 Tesla is the order of 1-10 $\%$ of typical exchange interactions. 
I note that intensities of external electromagnetic waves used in usual condensed-matter experiments 
are quite smaller than 1 [MV/cm] and 1 Tesla. Such a weak field is sufficient 
to observe linear responses of materials, but is not enough to do Floquet engineering.  
In the range of infrared and visible light, it is relatively easy to generate strong laser with intensity 
1.0-10 [MV/cm], while  such a strong laser in THz range is created only in the 
pulse form~\cite{Hirori11,Mukai16,Cavalleri17,Misawa13}. 
One should also note that (i) if we apply a strong CW beam with more than 10 [MV/cm] to crystals, 
most of them burn or evaporate, and (ii) the strength of external electric field usually become weaken 
in materials due to the relative permittivity. 

I summarize frequencies of electromagnetic wave and related physical quantities (electric and magnetic fields) 
in Tables~\ref{tab:MV} and \ref{tab:Units}.  
We have to carefully consider these values of electromagnetic waves,  
when we propose a realistic set up of Floquet engineering with a moderate value of $A/(\hbar\omega)$ 
(See Sec.~\ref{sec:3}). For example, if magnetic excitations are located in the range of 0.1-1 THz 
in a magnet considered, visible light is not suitable to perform the Floquet engineering 
through the spin-light coupling since $A/(\hbar\omega)$ (See Sec.~\ref{sec:2-4}) is too small. 
Instead, an intense laser with frequency of 2-3 THz would be better for the engineering.

Finally, I comment on wave length and spatial distribution of laser beams. 
Usually, the diffraction limit of electromagnetic wave is the order of its wave length. 
It is generally difficult to spatially focus laser beams beyond their diffraction length. 
For instance, 1THz wave has the wave length of $\sim$300[$\mu$m] which is much larger 
than lattice spacing of crystals. Therefore, 
AC electric or magnetic fields of low-frequency coherent waves can be viewed 
as spatially uniform AC fields for electrons in solids.
In other words, it is hard to introduce a microscopic spatial modulation in the THz range. 
However, in recent years, techniques based on meta-material and plasmonics~\cite{Pors12,Heeres14} have 
made possible to tightly focus AC fields beyond their diffraction limit. 
With these methods, it gradually becomes possible to create spatially modulated THz waves 
in micro- to nano-scales~\cite{Heeres14,Arikawa17,Fujita17}.

\section{Floquet Engineering in Magnets}\label{sec:4}
In this section, I review two proposals of Floquet engineering in magnetic insulators (quantum spin systems). 
Inverse Faraday effect in spin-orbit (SO) coupled electron systems~\cite{Pershan66}, 
dynamical localization~\cite{Dunlap86,Grossmann91,Iwai14,Lignier07}, and 
Floquet topological insulators~\cite{Oka09,Kitagawa11,Galitski11,Gedik13} 
are named as representative phenomena of Floquet engineering in solids.    
Particularly, the prediction of Floquet topological insulators have triggered the popularization of 
Floquet engineering in broad condensed-matter fields. 
Since this prediction, plenty of Floquet theories for electron or quasi-particle systems have been proposed.    

On the other hand, since the spin-light coupling is generally much weaker than the charge-light one, 
Floquet engineering in magnets had not been developed well.  
However, as I discussed in Sec.~\ref{sec:3}, THz laser science has strikingly grown and 
we can use intense THz laser pulses, which can be used to perform the Floquet engineering 
with spin-light couplings. 
I will explain theories for the inverse Faraday effect (ultrafast control of magnetization) with THz laser 
and ultrafast control of Dzyaloshinskii-Moriya (DM) interaction~\cite{Dzyaloshinskii58,Moriya60,Yoshida} 
in a class of multiferroic systems~\cite{Kimura03,KNB05,Tokura14}, respectively, 
in Sec.~\ref{sec:4-1} and \ref{sec:4-2}.

\subsection{Inverse Faraday Effect by THz Laser}\label{sec:4-1}
\begin{figure}[t]
\begin{center}
\includegraphics[scale=.5]{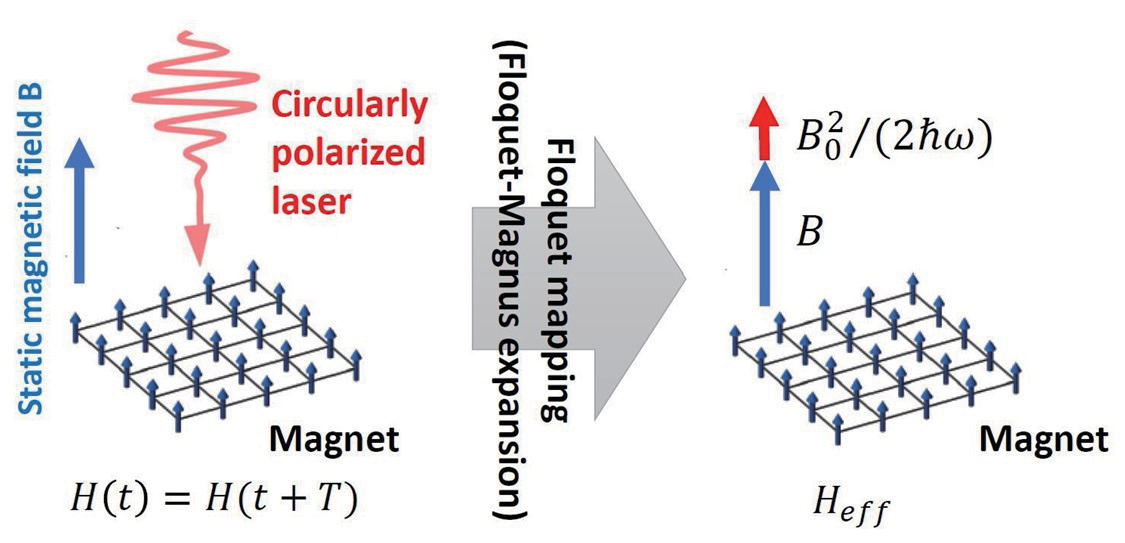}
\caption{Set up of our model (\ref{eq:Spin}) of a quantum spin system 
under a circularly polarized THz laser, and an image of Floquet mapping for the model.}
\label{fig:InverseFaraday}
\end{center}
\end{figure}

In this subsection, I consider a wide class of standard magnetic insulators 
(quantum spin systems)~\cite{Takayoshi14,Takayoshi14-2}. 
As I discussed in Sec.~\ref{sec:3}, magnetic excitations are distributed from 1 GHz to 10 THz, 
and thereby THz laser or wave are suitable for their Floquet engineering. 
I concentrate on a generic quantum spin model under the application of a static magnetic field 
and a circularly polarized wave, whose Hamiltonian (See Fig.~\ref{fig:InverseFaraday}) is given by 
\begin{align}
\label{eq:Spin}
\hat {\cal H}_{\rm mag}(t) &=  \hat H_{\rm mag}-{\bm B}\cdot \hat{\bm S}_{\rm tot} 
-{\bm B}(t)\cdot\hat{\bm S}_{\rm tot}. 
\end{align}
Here, $\bm B=(0,0,B)$ and $\bm B(t)=B_0(\sin(\omega t),-\cos(\omega t),0)$ respectively denote 
the Zeeman coupling constants of the static field and circularly polarized 
AC one with frequency $\omega$. 
They are defined as $B=g\mu_{\rm B} H_{\rm dc}$ and $B_0=g\mu_{\rm B} H_{\rm ac}$ 
($g$, $\mu_{\rm B}$, $H_{\rm dc}$ and $H_{\rm ac}$ are respectively electron g factor, 
Bohr magneton, and the strength of the static and AC magnetic fields), and 
${\bm S}_{\rm tot}=\sum_{\bm r}{\bm S}_{\bm r}$ is the total spin of the system. 
The AC field is in the $x$-$y$ plane and it means that the laser is irradiated from the $z$ direction. 
The first term $\hat H_{\rm mag}$ represents the static multiple-spin interaction 
and usually a Heisenberg-type exchange interaction is dominant there.   

For this driven system, only three Fourier components $\hat H_{0,+1,-1}$ are finite: 
$\hat H_0= \hat H_{\rm mag}-B\hat S^z_{\rm tot}$ and $\hat H_{\pm 1}=\mp\frac{i}{2} B_0\hat S^{\pm}_{\rm tot}$. 
Here, we define operators $\hat S^{\pm}_{\rm tot}=\hat S^{x}_{\rm tot}+\pm i\hat S^{y}_{\rm tot}$. 
The truncated Floquet Hamiltonian is thereby estimated as
\begin{align}
\label{eq:SpinFloquet}
\hat {H}_{\rm eff} &=  \hat H_{\rm mag}-\left(B+\frac{B_0^2}{2\hbar\omega}\right)\hat S_{\rm tot}^z
+{\cal O}\left((\hbar\omega)^{-2}\right). 
\end{align}
One finds that an effective Zeeman interaction $-(B_0^2/(2\hbar\omega))\hat S_{\rm tot}^z$ 
emerges owing to the applied laser. 
The Floquet Hamiltonian indicates that circularly polarized THz laser can change the value of magnetization 
$\langle{\hat S}^z_{\rm tot}\rangle$. An image of this Floquet theory is depicted in Fig.~\ref{fig:InverseFaraday}. 
It is well known that if we apply a circularly polarized 
high-frequency wave (such as visible light) to metallic systems with SO coupling, 
an effective Zeeman interaction also appears through the combination between charge-light and SO couplings. 
This Floquet engineering is called inverse Faraday effect~\cite{Pershan66} and 
it has been utilized in various studies of condensed-matter and applied physics~\cite{Kirilyuk10}. 
Equation~(\ref{eq:SpinFloquet}) shows that an inverse Faraday effect can also be generated 
even by applying circularly polarized THz (low-frequency) laser through the direct spin-light coupling. 

However, we should note that if the spin Hamiltonian $\hat H_{\rm mag}$ is SU(2)-symmetric 
(i.e., spin-rotation symmetric) like the Heisenberg model 
$\sum_{\langle{\bm r},{\bm r'}\rangle}\hat{\bm S}_{\bm r}\cdot\hat{\bm S}_{\bm r'}$, 
the static (time averaged) magnetization $\langle{\hat S}^z_{\rm tot}\rangle$ does not change 
even under the existence of the laser-driven Zeeman term~\cite{Takayoshi14}. 
This is understood by mapping the driven system to that on a rotating frame 
via the Unitary transform
\begin{align}
\label{eq:RotatingFrame}
\hat {\cal U}(t)&=\exp(i\hat S_{\rm tot}^z\omega t).
\end{align}
This mapping has been often used in the study of magnetic resonance~\cite{Slichter} and 
it eliminates the time-periodic Zeeman coupling since the Unitary rotation 
has the same frequency $\omega$ as the laser. 
After acting $\hat {\cal U}(t)$ to the Schr\o dinger equation from the left side, 
the transformed Hamiltonian $\hat H_{\rm u}=
\hat {\cal U}(t) \hat H(t)\hat {\cal U}^\dag(t)-i\hat {\cal U}(t) (\partial_t \hat {\cal U}^\dag(t))$ 
is given by
\begin{align}
\label{eq:RotatingFrameH}
\hat H_{\rm u}&=\hat H_{\rm mag}-B\hat S^z +B_0\hat S^y-\hbar\omega \hat S^z 
=\hat H_{\rm mag}-\bm B_{\rm u}\cdot \hat {\bm S},
\end{align}
where we have defined a new field $\bm B_{\rm u}=(0,-B_0,B+\hbar\omega)$. 
We note that the spin Hamiltonian $\hat H_{\rm mag}$ is invariant through the mapping 
due to the SU(2) symmetry. 
As expected, the in-plane AC field is mapped to a static field $B_0$ along the $S^y$ axis, 
while an additional Zeeman term $-\hbar\omega \hat S^z$ emerges. 
From this static model (\ref{eq:RotatingFrameH}), we see that 
the spin along the new field $\bm B_{\rm u}$ is conserved and therefore 
the static magnetization cannot grow even by applying any circularly polarized laser. 
For instance, a magnetic anisotropy such as single-ion terms $D_z\sum_{\bm r}({\hat S}^z_{\bm r})^2$ 
and Ising ones $\Delta_z\sum_{\langle{\bm r},{\bm r'}\rangle}{\hat S}^z_{\bm r}{\hat S}^z_{\bm r'}$ is 
necessary to break this conservation. Such anisotropies stem from SO coupling and 
after all it means that even the THz inverse Faraday effect requires 
an interaction connecting real and spin spaces like the usual high-frequency inverse Faraday effect. 
In Refs.~\cite{Takayoshi14,Takayoshi14-2}, the detailed condition for generating 
a large $\langle {\hat S}^z_{\rm tot}\rangle$ is discussed and its value is numerically computed.

\subsection{Ultrafast Control of Spin Chirality and Spin Current in Multiferroic Magnets}\label{sec:4-2}
In this subsection, we consider another type of magnetic insulators, a class of 
multiferroic magnets~\cite{KNB05,Tokura14}. 
Multiferroics stands for the system with cross-correlated multiple ferro-type orders in the broad sense. 
Particularly, in the last decade, researchers have actively studied a class of multiferroic magnetic insulators 
with a strong coupling (called magneto-electric (ME) coupling) between magnetic moments and 
electric polarization. Therefore, recently the word ``multiferroics'' have been used 
as this sort of magnets in a limited sense. In this subsection, I will use terminology ``multiferroics'' 
according to this convention. 
In such multiferroic magnets, there are several types of ME couplings and 
the electric polarization may be almost always approximated by a function of electron spins.

\begin{figure}[t]
\begin{center}
\includegraphics[scale=.45]{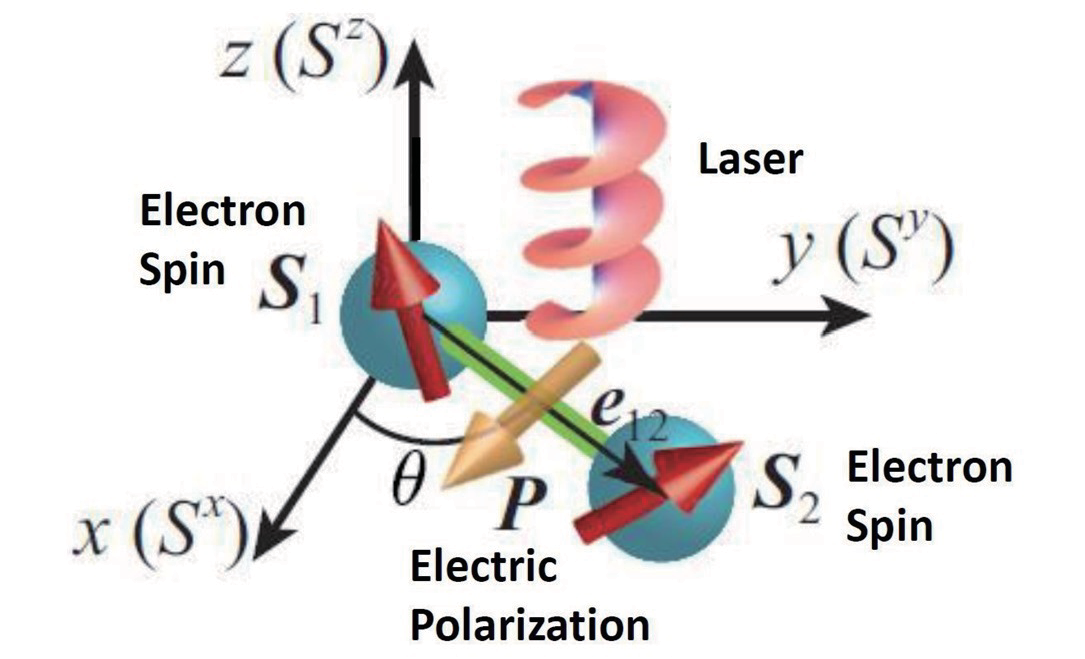}
\caption{Our two-spin multiferroic magnet under a circularly polarized THz laser. }
\label{fig:2SpinMultiferro}
\end{center}
\end{figure}

When we theoretically analyze the essential aspects of the Floquet engineering for multiferroics~\cite{Sato16}, 
it is enough to consider a simple two-spin multiferroic magnetic model described by Fig.~\ref{fig:2SpinMultiferro}. 
In this model, two spins resides on the $x$-$y$ plane and circularly polarized THz laser propagates 
parallel to the $z$ axis. The Hamiltonian is given by
\begin{align}
\label{eq:2Spin}
\hat {\cal H}_{\rm 2spin}(t) &=  \hat H_{\rm mag}-{\bm B}(t)\cdot\hat{\bm S}-{\bm E}(t)\cdot \hat{\bm P}
\end{align}
The first term $\hat H_{\rm mag}$ is the two-spin interaction like Eq.~(\ref{eq:Spin}). 
The second and third terms are driven by the circularly polarized laser: 
The second is the Zeeman interaction of the AC field 
$\bm B(t)=B_0(\sin(\omega t),-\cos(\omega t),0)$ with $B_0=g\mu_{\rm B} H_{\rm ac}$ 
and the third is the coupling between the AC electric field 
$\bm E(t)=E_0(\cos(\omega t), \sin(\omega t),0)$ and the electric polarization $\hat {\bm P}$. 
The symbol $\hat {\bm S}=\hat{\bm S}_1+\hat {\bm S}_2$ is the sum of two spins $\hat {\bm S}_{1,2}$, 
and the strength of magnetic field $H_{\rm ac}$ satisfies $H_{\rm ac}=E_0/c$ 
with $c$ being speed of light.

Let us compute the truncated Floquet Hamiltonian for the model~(\ref{eq:2Spin}). 
The Fourier components are given by $\hat H_0= \hat H_{\rm mag}$ and 
$\hat H_{\pm 1}=-\frac{1}{2}(E_0\hat P^{\pm}\pm i B_0\hat S^{\pm})$, where 
$\hat S^{\pm}=\hat S^x\pm i \hat S^y$ and $\hat P^{\pm}=\hat P^x\pm i \hat P^y$. 
Therefore, the effective Hamiltonian up to the $1/\omega$ order is 
\begin{align}
\label{eq:2SpinEff}
\hat H_{\rm eff} &=  \hat H_{\rm mag}-\frac{1}{2\hbar\omega}\Big[B_0^2 \hat S^z-iE_0^2 [\hat P^x,\hat P^y]
-iE_0B_0\left( [\hat P^x,\hat S^x]+ [\hat P^y,\hat S^y]\right)\Big].
\end{align}
This may be viewed as the generic formula for the Floquet Hamiltonian of multiferroic magnets 
under a circularly polarized laser~\cite{Sato16}. 
The terms proportional to $1/\omega$ are all the laser-driven interactions. 
The first term with $\hat S^z$ corresponds to the inverse Faraday effect discussed in the previous subsection.
The other terms proportional to $E_0B_0$ appear only in multiferroics with ME coupling, 
and they disappear in usual magnetic insulators (See Sec.~\ref{sec:4-1}).

With this formula of Eq.~(\ref{eq:2SpinEff}), let us consider a multiferroic model with a concrete ME coupling.  
Among several ME couplings, I here focus on so-called inverse DM type coupling, 
in which the electric polarization is given by the outer product of two neighboring spins 
$\hat {\bm V}=\hat{\bm S}_1\times\hat {\bm S}_2$ (vector spin chirality)~\cite{KNB05}:
\begin{align}
\label{eq:KNB}
\hat {\bm P} &=g_{\rm me}\bm e_{12}\times (\hat{\bm S}_1\times\hat {\bm S}_2), 
\end{align}
where $g_{\rm me}$ is the ME coupling constant and generally depends on the frequency $\omega$. 
The vector $\bm e_{12}=(\cos\theta,\sin\theta,0)$ is the unit vector connecting two spins as shown in 
Fig.~\ref{fig:2SpinMultiferro} and the symbol $\times$ denotes outer product. 
This inverse DM type coupling is originated from SO coupling and is known to often emerge 
in multiferroics with super-exchanges between a transition metal ion and an oxygen ion 
(such as Mn oxides and Cu oxides)~\cite{Kimura03,KNB05,Tokura14,Cheong09,Furukawa10}.  
Substituting the polarization $\hat {\bm P}=
g_{\rm me}(\sin\theta\hat{ V}^z,-\cos\theta\hat V^z,\cos\theta\hat V^y-\sin\theta\hat V^x)$ 
to the Floquet Hamiltonian (\ref{eq:2SpinEff}), we obtain 
\begin{align}
\label{eq:KNBeff}
\hat H_{\rm eff} &= \hat H_{\rm mag}-\frac{B_0^2}{2\hbar\omega}\hat S^z
-\frac{g_{\rm me}E_0B_0}{2\hbar\omega}\bm e_{12}\cdot\hat {\bm V}. 
\end{align}
The final term is generated by the cross-correlation between $E_0$ and $B_0$, and 
may be called a laser-driven DM interaction. 
DM interactions generally make two neighboring spins perpendicularly oriented with each other, 
while standard exchange interactions prefer a collinear spin structure (parallel or anti-parallel). 
Therefore, the co-existence of exchange and DM interactions usually creates a non-collinear spin structure, 
in which neighboring spins take a certain angle $\phi\neq 0,\pi$. 
The effective model (\ref{eq:KNBeff}) thus indicates that if we apply a circularly polarized laser 
to a multiferroic magnet with inverse DM coupling, a non-collinear magnetic structure can be created or annihilated.

\begin{figure}[t]
\begin{center}
\includegraphics[scale=.5]{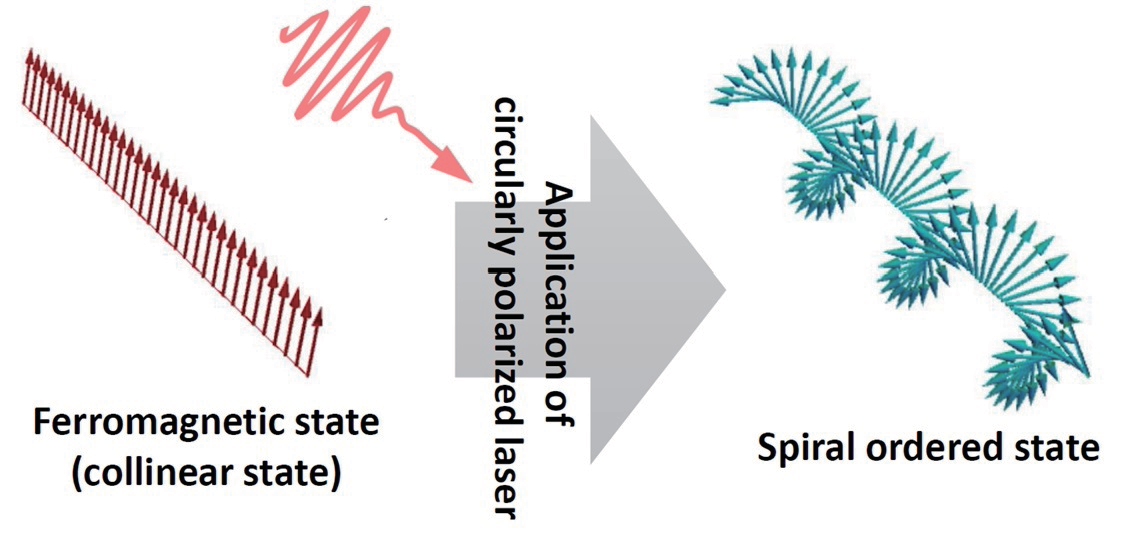}
\caption{Floquet engineering of a multiferroic chain with inverse DM coupling under a circularly polarized THz laser. }
\label{fig:MultiferroChain}
\end{center}
\end{figure}

It is straightforward to extend the result of Eq.~(\ref{eq:KNBeff}) to many-spin multiferroic models. 
For instance, an one-dimensional (1D) multiferroic model along the $x$ direction 
under a circularly polarized laser is described by the following Hamiltonian 
\begin{align}
\label{eq:1D}
\hat {\cal H}_{\rm 1D} &= \sum_j J\hat{\bm S}_j\cdot\hat{\bm S}_{j+1}-{\bm B}\cdot\hat{\bm S}_{\rm tot}
-{\bm B}(t)\cdot\hat{\bm S}_{\rm tot}-{\bm E}(t)\cdot \hat{\bm P}_{\rm tot},
\end{align}
where $\bm S_j$ is the electron spin on $j$th site, ${\bm S}_{\rm tot}=\sum_j {\bm S}_j$ is the total spin, 
and the polarization is given by 
$\hat{\bm P}_{\rm tot}=g_{\rm em}\sum_j {\bm e}\times(\hat{\bm S}_j\times\hat {\bm S}_{j+1})$ 
with ${\bm e}=(1,0,0)$ being the unit vector along the chain ($x$) direction. 
The first term is the 1D exchange interaction between neighboring spins and 
the second is the static Zeeman interaction. 
The third and fourth terms are generated by applied laser fields $\bm B(t)$ and $\bm E(t)$.  
After some algebra, we arrive at the Floquet Hamiltonian
\begin{align}
\label{eq:1Deff}
\hat H_{\rm eff} &= \sum_j J\hat{\bm S}_j\cdot\hat{\bm S}_{j+1}
-{\bm B}\cdot\hat{\bm S}_{\rm tot}
-\frac{B_0^2}{2\hbar\omega}\hat S_{\rm tot}^z
-\frac{g_{\rm me}E_0B_0}{2\hbar\omega}\sum_j(\hat{\bm S}_j\times\hat {\bm S}_{j+1})^x. 
\end{align}
The final term is the laser-driven uniform DM interaction which prefers a spiral spin structure. 
A finite spin chirality $\langle V_{\rm tot}^x\rangle
=\sum_j\langle(\hat{\bm S}_j\times\hat {\bm S}_{j+1})^x\rangle$ is shown~\cite{Sato16} to be created 
when we numerically solve the Schr\o dinger equation for the driven multiferroic model (\ref{eq:1D}). 
This is owing to the competition between exchange and laser-driven DM interactions in Eq.~(\ref{eq:1Deff}). 
An image of this Floquet engineering is depicted in Fig.~\ref{fig:MultiferroChain}.  

A local spin chirality $\langle V^x_j\rangle=\langle(\hat{\bm S}_j\times\hat {\bm S}_{j+1})^x\rangle$ 
can be viewed as a local spin current~\cite{KNB05,SpinCurrent12}. 
From the equation of continuity of spin, one sees that a local spin flow can appear 
if neighboring spin chiralities have different expectation values: 
$\langle V^x_j\rangle-\langle V^x_{j+1}\rangle\neq 0$. 
Such a spin current is numerically shown~\cite{Sato16} to be created 
if we apply a ``spatially modulated'' laser to the 1D multiferroic model. 
As I mentioned in Sec.~\ref{sec:3}, spatially modulated THz laser could be generated 
with meta-material techniques~\cite{Pors12,Heeres14,Arikawa17}.

In addition to inverse DM interaction, the magneto-striction mechanism is also famous as 
a representative of ways of generating ME couplings~\cite{Tokura14}. 
This mechanism usually stems from spin-phonon coupling and leads to a coupling between 
a local electric polarization and a local exchange energy $\hat {\bm S}_{\bm r}\cdot\hat {\bm S}_{\bm r'}$. 
Of course, it is generally possible to propose a Floquet engineering with such a magneto-striction ME coupling. 
In fact, it has been predicted~\cite{Sato14} that a topological spin liquid can be created 
if we apply a circularly polarized laser to a honeycomb Kitaev model~\cite{Kitaev,Jackeli1,Jackeli2} 
with a striction type ME coupling.

\section{Summary and Outlook}\label{sec:5}
In this Chapter, I have reviewed the basic theoretical part of 
Floquet engineering~\cite{Bukov,Eckardt15,Eckardt17,Oka19}, focusing on 
isolated periodically-driven quantum systems. 
Floquet theorem, Floquet-Magnus expansion for deriving the effective Floquet Hamiltonian, and 
its physical meanings have been explained in Sec.~\ref{sec:2}. 
The truncated Floquet Hamiltonian can be used to correctly describe the short-time behavior of driven systems. 
Then, in Sec.~\ref{sec:3}, I have discussed some information about currently-available laser and 
electromagnetic waves which is important to perform Floquet engineering in materials, especially, solids. 
The intensity of 1 [MV/cm] gives a reference amplitude of AC electric field 
for effective engineering of physical quantities. 
Finally, I have explained two examples of Floquet engineering in 
magnetic insulators: inverse Faraday effect with THz laser 
in generic magnetic insulators~\cite{Takayoshi14,Takayoshi14-2} 
and ultrafast control of spin chirality in multiferroics~\cite{Sato16}.

Further development of Floquet theory is necessary 
to more accurately predict Floquet engineered quantities in materials. 
An important research direction is the development of sophisticated theories 
treating effect of environment (i.e., dissipation) 
because dissipation effect cannot be ignored in real materials driven by laser. 
Approaches based on non-equilibrium Green's function~\cite{Haug,Aoki} and 
quantum master equation~\cite{Breuer,Breuer00,Ikeda19-2} 
have high potential to provide a deep insight to dissipative periodically driven systems.  
These studies would also contribute to the development of the fundamental of non-equilibrium physics.

\section*{Acknowledgments}
I would like to thank all the collaborators of our recent works for laser-driven phenomena, 
especially, Shintaro Takayoshi, Takashi Oka, Tatsuhiko N. Ikeda, Hiroaki Ishizuka, Horoyuki Fujita, and 
Sho Higashikawa. I was supported by JSPS KAKENHI (Grant No. 17K05513 and No. 20H01830) 
and a Grant-in-Aid for Scientific Research on Innovative Areas
``Quantum Liquid Crystals'' (Grant No. JP19H05825). 



\end{document}